\documentclass[twocolumn]{aastex631}

\usepackage{threeparttable}
\usepackage{subfigure}
\usepackage{longtable}
\begin{document}
\title{The timing and spectral properties of the 2022 outburst of SGR J1935+2154 observed with NICER}

\author[0009-0005-2228-0618]{Fu Yu-Cong}
\affiliation{Institute for Frontiers in Astronomy and Astrophysics, Beijing Normal University, Beijing, 102206, China}
\affiliation{School of Physics and Astronomy, Beijing Normal University, Beijing, 100875, China}

\author[0000-0002-0633-5325]{Lin Lin}
\affiliation{Institute for Frontiers in Astronomy and Astrophysics, Beijing Normal University, Beijing, 102206, China}
\affiliation{School of Physics and Astronomy, Beijing Normal University, Beijing, 100875, China}

\author{Ge Ming-Yu}
\affiliation{Key Laboratory of Particle Astrophysics, Institute of High Energy Physics, Chinese Academy of Sciences, Beijing, 100049, China}

\author{Enoto Teruaki}
\affiliation{Department of Physics, Graduate School of Science, Kyoto University, Kyoto, 606-8502, Japan}

\author{Hu Chin-Ping}
\affiliation{Department of Physics, National Changhua University of Education, Changhua City, Taiwan}

\author{Younes George}
\affiliation{Astrophysics Science Division, NASA Goddard Space Flight Center, 8800 Greenbelt Road, Greenbelt, MD, 20771, USA}
\affiliation{Center for Space Sciences and Technology, UMBC, Baltimore, MD, 21250, USA}

\author{Göğüş Ersin}
\affiliation{Sabancı University, Faculty of Engineering and Natural Sciences, İstanbul, 34956, Turkey}

\author{Malacaria Christian}
\affiliation{INAF-Osservatorio Astronomico di Roma, Via Frascati 33, Monteporzio Catone (RM), I-00078, Italy}

\correspondingauthor{Lin Lin}
\email{llin@bnu.edu.cn}



\begin{abstract}
The magnetar SGR J1935+2154 entered a new active episode on October 10, 2022, with X-ray bursts and enhanced persistent emission.
At the tail of high burst rate interval, lasting several hours, radio bursts were detected, revealing the connection between the X-ray activities and radio emissions.
We analyzed observations of SGR J1935+2154 for nearly three months, using data from Neutron Star Interior Composition Explorer (NICER).
We report the timing and spectral results following the onset of this outburst.
In general, the X-ray flux of the persistent emission decays exponentially. 
While a flare is evident on the light curve, a fast radio burst (FRB) was detected immediately following the peak of this flare.
We found a phase jump of pulse profile, with a deviation of $0.16\pm0.03$ phase, which is related to the glitch. 
The spectra are well fit with the combination of a blackbody and a power law model. The decay of the outburst is dominated by the drop of the non-thermal component, which also leads to the increase of thermal proportion. 
The photon index of the power law is inversely correlated with both the unabsorbed flux and the burst rate.
We find that unlike the large variety of the persistent emission around FRB 221014, the X-ray properties are very stable when FRBs 221021 and 221201 happened.  
These results manifest the connection between glitch, phase jump, X-ray burst, and radio burst, crucial for studying the mutation in twisted magnetic fields and constraining the trigger mechanism of radio bursts.
\end{abstract}



\section{Introduction} \label{sec:intro}
Magnetars are a special type of isolated neutron stars (NSs), with distinctive phenomenons in X-ray and radio bands, which are mainly powered by their immense (up to $\sim 10^{15}$ G) magnetic fields \citep[e.g.,][]{1998Natur.393..235K, 2014ApJS..212....6O, 2017ARA&A..55..261K}.
During the active period, the brightening of X-ray emission from magnetars is often accompanied by bursts.
The bursts can occur sporadically with a few bursts, or massive bursts are concentrated within a few hours \citep[e.g.,][]{2008ApJ...685.1114I, 2011ApJ...740L..16L, 2015ApJ...807...93A, 2022ApJS..260...24C}.
The persistent emission level also increases rapidly, accompanied by the variability of spectral and timing properties \citep[e.g.,][]{2004ApJ...605..378W, 2017ApJ...851...17Y, 2022MNRAS.516..602B}.
Discovered for the first time in 2007 \citep{2007Sci...318..777L}, Fast radio bursts (FRBs) are short pulses observed in the radio sky, and the explanation of their origin has become a central unresolved problem in astronomy \citep{2019ARA&A..57..417C, 2019A&ARv..27....4P}.
Although the physical mechanism of FRBs is not yet clear \citep{2020Natur.587...45Z, 2021SCPMA..6449501X}, the connection between the magnetars and the origin of some FRBs has been established \citep{2019A&ARv..27....4P, 2020Natur.587...59B, 2020Natur.587...63L, 2023ApJ...953...67G, 2024RAA....24a5016G}.

SGR J1935+2154 was discovered during its outburst in 2014 \citep{2014GCN.16520....1S}.
Then, the spin period of NS was calculated as about 3.24 s with a spin-down rate of $\dot{P}\rm \sim1.4\times10^{-11} s\,s^{-1}$, which implies a surface dipolar magnetic field strength of $B\sim 2.2 \times 10^{14}$ G \citep{2016MNRAS.457.3448I}.
Since then, as one of the most active magnetars, SGR J1935+2154 has shown outbursts in 2014, 2015, 2016, 2019, and 2020, which exhibit the different properties of persistent emission of the source \citep[e.g.,][]{2017ApJ...847...85Y, 2020ApJ...902L..43L, 2020ApJ...893..156L, 2020ApJ...904L..21Y, 2022MNRAS.516..602B}.  
The X-ray spectra below 10 keV are described well with a blackbody+powerlaw (BB+PL) or BB+BB model, and the cool BB temperature $kT_{\rm BB}$ is about 0.47 keV \citep{2017ApJ...847...85Y}.
In 2020 outburst, with the decay in the luminosity, the contribution of the PL component was observed to decrease from $\sim$ 75\% to $\sim$ 45\% \citep{2022MNRAS.516..602B, 2017ApJ...847...85Y}.
X-ray burst and FRB 200428 are both observed in 2020 outburst, suggesting that they originate from the same source \citep{2020Natur.587...54C, 2020Natur.587...59B, 2021NatAs...5..378L, 2021NatAs...5..372R}.

In 2022 October, SGR J1935+2154 entered a new active period, with several instruments triggered on the bursts \citep{2022ATel15667....1P, 2022ATel15672....1R, 2022ATel15698....1L, 2022ATel15690....1E}.
During the peak of the active period, many bursts tend to cluster together, resulting in a high burst rate interval \citep{2022ATel15674....1Y}.
After this interval, several FRB-like bursts were detected by GBT \citep{2022ATel15697....1M}, CHIME \citep{2022ATel15681....1D, 2022ATel15792....1P}, and Yunnan 40m radio telescope \citep{2022ATel15707....1H}.
The details of the FRBs are outlined as follows:
\begin{itemize}
\item FRB 221014, detected by CHIME/FRB on October 14, 2022, at 19:21:39.130 UTC ($\sim$ MJD 59866.82), with a fluence of $9.7 \pm 6.7$ kJy ms \citep{2022ATel15681....1D, 2023arXiv231016932G}, was accompanied by X-ray bursts captured by Konus-Wind \citep{2022ATel15686....1F} and GECAM \citep{2022ATel15682....1W}.
\item FRB 221021, detected by Yunnan 40m radio telescope on October 21, 2022, at 10:01:45.84215 UT ($\sim$ MJD 59873.42, \citealp{2022ATel15707....1H}), was accompanied by an X-ray burst captured by \textit{Insight}-HXMT \citep{2022ATel15708....1L, 2022ATel15714....1L}.
\item FRB 221201, was detected by CHIME/FRB on December 1, 2022, at 22:06:59.0762 UTC ($\sim$ MJD 59914.92), with a fluence of $23.7 \pm 18.0$ kJy ms \citep{2022ATel15792....1P, 2023arXiv231016932G}.
\end{itemize}

During the 2022 outburst, the spectrum of the persistent emission observed with XMM-Newton and NuSTAR is also well described by the BB+PL model below $\sim$ ~25 keV, and there are no significant changes in the BB temperature ($kT_{\rm BB}$ $\sim$ 0.4 keV) between the two epochs on October 15$-$18 and October 22, 2022 \citep{2024ApJ...965...87I}.
\citet{2024Natur.626..500H} analyzed the data observed with NICER and NuSTAR during a month following the outburst, and discovered an unprecedented double glitch\footnote{The glitch means a sudden spin-up of the NS.} associated with FRB 221014.

In this study, using NICER data of SGR J1935+2154 during its 2022 outburst, spanning nearly three months, we report the long-term evolution of the timing and spectral properties of this source.   
The observations and data reduction are presented in Section \ref{sec:Observations}, the data analysis and results are described in Section \ref{sec:ANALYSIS AND RESULTS}, and, finally, the discussions and conclusions are given in Section \ref{sec:Discussion}. 

\section{Observations and Data Reduction}
\label{sec:Observations}
\begin{figure*}
    \centering
    \includegraphics[width=0.98\textwidth]{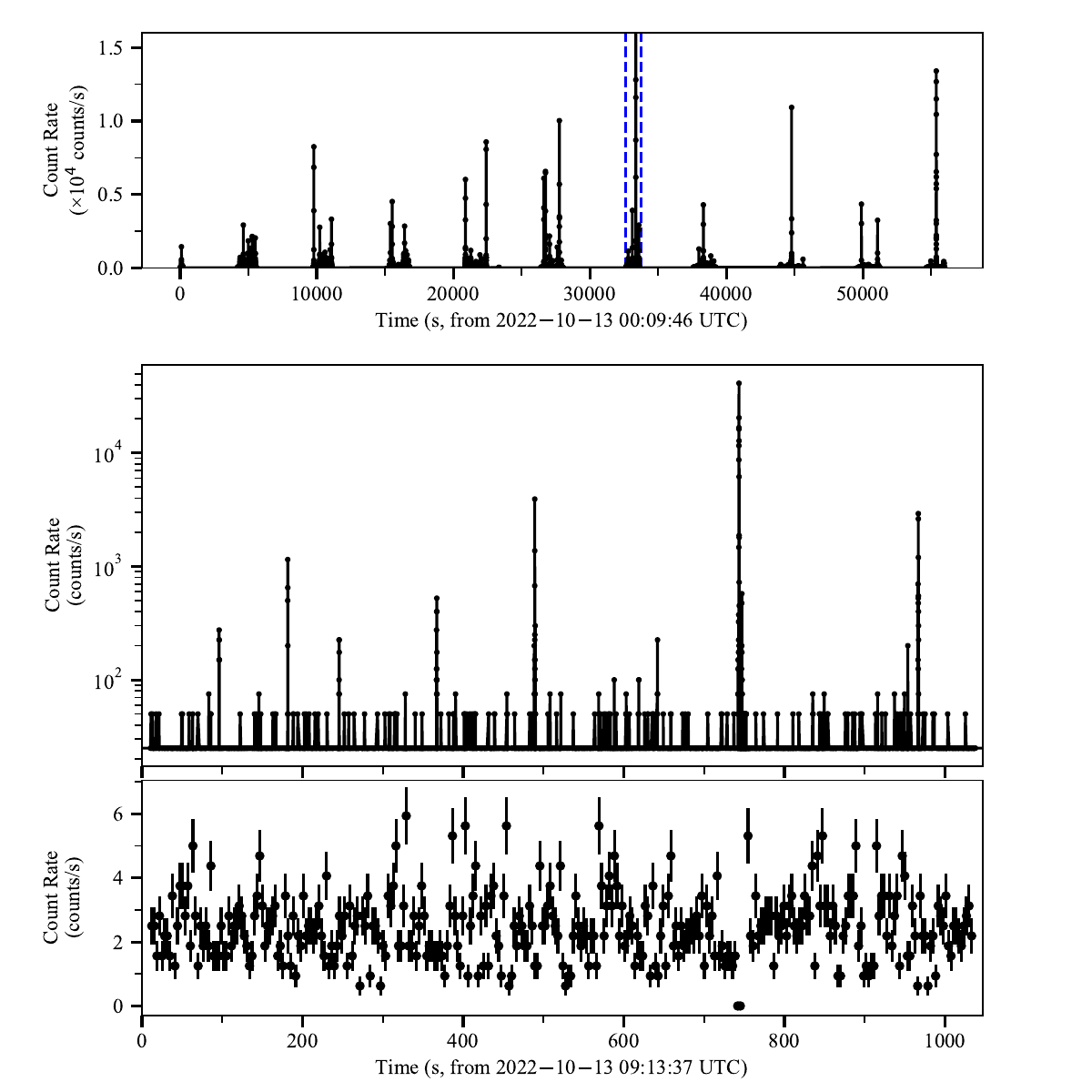}
    \caption{NICER light curve in the 0.3--12 keV energy range.
    Top panel: the light curve of Obs. ID 5020560107 with a 4 ms resolution. The GTI between two blue dashed vertical lines is enlarged and displayed in the middle panel.
    Middle panel: the zoomed-in light curve of a GTI with a 4 ms resolution.
    Bottom panel: the zoomed-in light curve with a 3.2 s resolution is calculated after removing all the identified bursts.
    }
    \label{fig:lc-1}
\end{figure*}

NICER is an X-ray timing and spectral instrument, which is successfully launched in 2017 and installed on the International Space Station \citep{2017NatAs...1..895G}. 
NICER consists of 56 X-ray concentrator optics, 52 of which are currently operating. It covers the 0.2--12 keV energy range and provides a collecting area of 1900 $\rm cm^2$ at 1.5 keV \citep{2016SPIE.9905E..1HG}.

NICER started monitoring SGR J1935+2154 on October 12, 2022, at 17:32:40 UTC, during its 2022 active episode, with the detection of over 100 bursts and enhanced persistent emission \citep{2022ATel15674....1Y}.
In this work, we analyzed the data from October 12 to December 21, 2022, with a total of 38 observations, as shown in Table \ref{tab:table3}. 
Each observation consists of several good time intervals (GTIs).
For observations with high emission levels within the first three days, we divide them into smaller segments based on GTIs for further analysis.

We processed the NICER observations using NICERDAS (v.10) as part of HEASOFT (v.6.31.1), along with the calibration database (CALDB) version 20221001.
The standard calibration and filtering processes are performed by the task \texttt{nicerl2}\footnote{\href{https://heasarc.gsfc.nasa.gov/lheasoft/ftools/headas/nicerl2.html}{https://heasarc.gsfc.nasa.gov/lheasoft/ftools/headas/nicerl2.html}} with the default parameters.
The spectra, background, and associated responses in the NICER-recommended way\footnote{\href{https://heasarc.gsfc.nasa.gov/docs/nicer/analysis_threads/cal-recommend/}{https://heasarc.gsfc.nasa.gov/docs/nicer/analysis\_threads/cal-recommend/}} are generated by the task \texttt{nicerl3-spect}\footnote{\href{https://heasarc.gsfc.nasa.gov/lheasoft/ftools/headas/nicerl3-spect.html}{https://heasarc.gsfc.nasa.gov/lheasoft/ftools/headas/nicerl3-spect.html}} with the \texttt{SCORPEON} background model.
The light curves and associated background estimation are generated using a pipeline task \texttt{nicerl3-lc}\footnote{\href{https://heasarc.gsfc.nasa.gov/docs/nicer/analysis_threads/nicerl3-lc/}{https://heasarc.gsfc.nasa.gov/docs/nicer/analysis\_threads/nicerl3-lc/}} with the \texttt{Space Weather} background model.

To study the evolution of the persistent emission, we excluded all the identified bursts. 
We apply a Poissonian procedure to identify the bursts \citep[e.g.,][]{2004ApJ...607..959G} with a significance higher than 3-$\sigma$, as described in \citet{2020ApJ...904L..21Y}.
To avoid interference from burst tails, we carefully remove the nearby bases of the bursts as much as possible, ensuring a clear analysis of the persistent emission.
Figure \ref{fig:lc-1} shows an example of burst subtraction, where the initial light curve of Obs. ID 5020560107 with a 4 ms resolution is displayed in the top panel, which is divided into 11 GTIs. 
The middle panel of Figure \ref{fig:lc-1} presents a zoomed-in view of the GTI between two blue dashed vertical lines.
After removing all the identified bursts, the light curve with 3.2 s resolution is shown in the bottom panel.

The events in the energy range 0.8--4 keV after the barycenter correction are used for timing analysis, because the effective area within this range is larger.
We use the epoch-folding technique \citep{1987A&A...180..275L} to search the initial period of the neutron star \citep[e.g.,][]{2020ApJ...902L...2B, 2020JHEAp..27...38T, 2023MNRAS.521..893F}, and use the TEMPO2 \citep{2006MNRAS.372.1549E} version 2022.05.1 to update the more accurate ephemeris of the source.
Times of arrival (ToAs) are acquired through $Z_1^2$ searching, and the minimum phase in each profile is used as the TOA for that specific observation \citep{2012ApJS..199...32G, 2019NatAs...3.1122G, 2020ApJ...904L..21Y}.
For the spectral analysis, we fit all the spectra using XSPEC \citep{1996ASPC..101...17A} version 12.13.0c.
A photo-electric absorption model \texttt{wabs} \citep{1982GeCoA..46.2363A, 1983ApJ...270..119M} is used to calculate the interstellar absorption of SGR J1935+2154.
The errors of the parameters are given at the level of 1-$\sigma$ uncertainty.

\begin{longtable*}{lccccccc}

\caption{NICER observations of SGR J1935+2154 analyzed in this work.  \label{tab:table3}}\\  

\hline  
\hline 
\multicolumn{1}{l}{Obs. ID-GTI$\rm ^{a}$} &   
\multicolumn{1}{c}{Time$\rm ^{b}$ (MJD)} &   
\multicolumn{1}{c}{Exposure (s)} &   
\multicolumn{1}{c}{Burst$\rm ^{c}$} &   
\multicolumn{1}{c}{Count Rate$\rm ^{d}$} &   
\multicolumn{1}{c}{Index} &   
\multicolumn{1}{c}{Flux$\rm ^{e}$} &   
\multicolumn{1}{c}{Part $\rm ^{f}$}\\  
\hline  
\endfirsthead 
\multicolumn{8}{c}{\tablename\ \thetable\ --Continued from previous page} \\  
\hline  
\hline  
\multicolumn{1}{l}{Obs. ID-GTI$\rm ^{a}$} &   
\multicolumn{1}{c}{Time$\rm ^{b}$ (MJD) } &   
\multicolumn{1}{c}{Exposure (s)} &   
\multicolumn{1}{c}{Burst$\rm ^{c}$} &   
\multicolumn{1}{c}{Count Rate$\rm ^{d}$} &   
\multicolumn{1}{c}{Index} &   
\multicolumn{1}{c}{Flux$\rm ^{e}$} &   
\multicolumn{1}{c}{Part $\rm ^{f}$}\\  
\hline  
\endhead
\multicolumn{8}{c}{{Continued on next page}} \\ \hline  
\endfoot  
\endlastfoot
5020560106-01&59864.7364 &532.97	&5  &$2.15\pm0.08$ &$	2.42\pm	0.12$&$	2.30\pm	0.12$& I\\
5020560106-02&59864.8011 &655.96	&2  &$2.27\pm0.06$ &$	2.31\pm	0.13$&$	1.75\pm	0.14$& I\\
5020560106-03&59864.8667 &75.00	    &0  &$2.03\pm0.16$ &$	2.40\pm	0.61$&$	1.70\pm	0.44$& I\\
5020560107-01&59865.0068 &212.98	&2  &$2.48\pm0.11$ &$	2.25\pm	0.27$&$	1.62\pm	0.18$& I\\
5020560107-02&59865.0555 &1320.92   &11 &$2.71\pm0.05$ &$	2.63\pm	0.08$&$	2.10\pm	0.06$& I\\
5020560107-03&59865.1193 &1391.92   &6  &$2.68\pm0.05$ &$	2.29\pm	0.08$&$	2.05\pm	0.06$& I\\
5020560107-04&59865.1838 &1515.91   &9  &$2.62\pm0.04$ &$	2.59\pm	0.07$&$	2.03\pm	0.06$& I\\
5020560107-05&59865.2482 &1574.90   &7  &$2.29\pm0.04$ &$	2.63\pm	0.09$&$	1.77\pm	0.05$& I\\
5020560107-06&59865.3128 &1581.90   &7  &$2.37\pm0.04$ &$	2.68\pm	0.08$&$	1.87\pm	0.06$& I\\
5020560107-07&59865.3846 &1027.93   &7  &$2.29\pm0.05$ &$	2.49\pm	0.10$&$	1.75\pm	0.07$& I\\
5020560107-08&59865.4419 &1631.90   &6  &$2.23\pm0.04$ &$	2.78\pm	0.08$&$	1.68\pm	0.05$& I\\
5020560107-09&59865.5145 &1764.86   &2  &$1.71\pm0.03$ &$	2.59\pm	0.09$&$	1.31\pm	0.04$& I\\
5020560107-10&59865.5814 &1569.87   &2  &$1.86\pm0.04$ &$	2.55\pm	0.10$&$	1.30\pm	0.05$& I\\
5020560107-11&59865.6396 &1310.91   &3  &$1.98\pm0.04$ &$	2.77\pm	0.10$&$	1.56\pm	0.06$& I\\
5576010101-01&59865.7016 &1951.87   &2  &$2.00\pm0.03$ &$	2.74\pm	0.09$&$	1.17\pm	0.04$& I\\
5576010101-02&59865.7676 &1160.93   &2  &$1.97\pm0.04$ &$	2.59\pm	0.13$&$	1.14\pm	0.06$& I\\
5576010102-01&59866.1514 &1892.87   &10 &$2.41\pm0.04$ &$	2.74\pm	0.09$&$	1.17\pm	0.04$& I\\
5576010102-02&59866.2160 &1889.87   &5  &$2.67\pm0.04$ &$	2.58\pm	0.08$&$	1.79\pm	0.26$& I\\
5576010102-03&59866.4094 &1289.92   &6  &$5.14\pm0.06$ &$	1.79\pm	0.10$&$	3.06\pm	0.35$& I\\
5576010102-04&59866.4813 &1303.90   &10 &$5.54\pm0.07$ &$	2.07\pm	0.06$&$	4.35\pm	0.10$& I\\
5576010102-05&59866.6097 &877.94	&1  &$3.87\pm0.07$ &$	2.15\pm	0.12$&$	1.95\pm	0.08$& I\\
\hline
5576010102-06&59866.7331 &1535.90   &36 &$12.28\pm0.09$&$   1.31\pm	0.02$&$	19.29\pm0.18$& II\\
5576010102-07&59866.8676 &807.95	&5  &$6.02\pm0.09$ &$	2.01\pm	0.06$&$	4.97\pm	0.13$& II\\
5576010102-08&59866.9279 &378.98	&0  &$6.77\pm0.13$ &$	1.96\pm	0.12$&$	4.19\pm	0.20$& II\\
5576010103-01&59867.0008 &867.93	&1  &$4.83\pm0.07$ &$	2.14\pm	0.11$&$	3.73\pm	0.17$& II\\
5576010103-02&59867.0680 &1311.89   &0  &$5.97\pm0.07$ &$	2.07\pm	0.10$&$	3.59\pm	0.15$& II\\
5576010103-03&59867.1835 &1973.87   &0  &$4.12\pm0.05$ &$	2.37\pm	0.08$&$	2.81\pm	0.09$& II\\
5576010103-04&59867.3236 &998.92	&0  &$3.98\pm0.06$ &$	2.24\pm	0.14$&$	2.37\pm	0.14$& II\\
5576010103-05&59867.3880 &999.92	&0  &$3.24\pm0.06$ &$	2.56\pm	0.14$&$	2.20\pm	0.10$& II\\
5576010103-06&59867.5167 &1019.92   &0  &$2.40\pm0.06$ &$	2.32\pm	0.14$&$	2.03\pm	0.12$& II\\
5576010103-07&59867.7805 &584.95	&0  &$1.41\pm0.06$ &$	2.49\pm	0.21$&$	1.13\pm	0.15$& II\\
5576010104   &59868.0968 &	9993 	&2	&$1.99\pm0.02$ &$   2.62\pm 0.03$&$ 1.80\pm 0.04$& II\\
5576010105	 &59869.0539 &	2609 	&4	&$1.83\pm0.03$ &$   2.61\pm 0.07$&$ 1.94\pm 0.05$& II\\
5576010106	 &59870.1655 &	2474 	&0	&$1.28\pm0.03$ &$   2.97\pm 0.06$&$ 1.43\pm 0.05$& II\\
5576010108	 &59872.2922 &	1693 	&0	&$1.23\pm0.04$ &$   3.22\pm 0.10$&$ 1.27\pm 0.07$& II\\
5576010109	 &59873.0013 &	3021 	&0	&$0.98\pm0.03$ &$   3.05\pm 0.20$&$ 0.92\pm 0.06$& II\\
\hline
5576010110	 &59874.0871 &	2034 	&1	&$1.46\pm0.04$ &$   2.88\pm 0.11$&$ 1.38\pm 0.05$& III\\
5020560108	 &59875.9127 &	260 	&0	&$2.07\pm0.22$ &$   2.39\pm 0.84$&$ 0.76\pm 0.18$& III\\
5020560109	 &59877.2640 &	670 	&0	&$0.99\pm0.07$ &$   3.04\pm 0.18$&$ 0.89\pm 0.06$& III\\
5576010111	 &59878.3503 &	6596 	&0	&$1.04\pm0.02$ &$   2.90\pm 0.06$&$ 1.17\pm 0.03$& III\\
5576010112	 &59886.5579 &	753 	&0	&$1.02\pm0.08$ &$   2.82\pm 0.28$&$ 1.24\pm 0.18$& III\\
5020560110	 &59888.3810 &	1194 	&0	&$0.96\pm0.04$ &$   3.55\pm 0.24$&$ 0.80\pm 0.06$& III\\
5020560111	 &59889.0319 &	728 	&0	&$0.89\pm0.05$ &$   2.81\pm 0.19$&$ 0.76\pm 0.05$& III\\
5576010113	 &59889.3457 &	1094 	&0	&$0.90\pm0.03$ &$   3.21\pm 0.17$&$ 0.77\pm 0.03$& III\\
5020560112	 &59892.9453 &	1161 	&2	&$1.03\pm0.04$ &$   2.68\pm 0.12$&$ 1.38\pm 0.05$& III\\ 
5576010114	 &59893.0096 &	1134 	&0	&$1.27\pm0.05$ &$   2.82\pm 0.13$&$ 1.19\pm 0.06$& III\\ 
5020560113	 &59893.9125 &	1717 	&1	&$1.04\pm0.03$ &$   3.35\pm 0.15$&$ 1.06\pm 0.04$& III\\ 
5020560114	 &59894.1060 &	8347 	&1	&$1.41\pm0.13$ &$   3.01\pm 0.05$&$ 1.38\pm 0.02$& III\\
5020560115	 &59895.0090 &	10778 	&2	&$1.42\pm0.07$ &$   2.72\pm 0.03$&$ 1.28\pm 0.03$& III\\
5576010115	 &59895.7272 &	437 	&0	&$0.89\pm0.05$ &$   3.03\pm 0.51$&$ 0.78\pm 0.12$& III\\
5020560116	 &59896.0497 &	7068 	&0	&$0.92\pm0.03$ &$   3.08\pm 0.05$&$ 1.24\pm 0.03$& III\\
5020560117	 &59897.0086 &	10996 	&2	&$0.87\pm0.02$ &$   3.02\pm 0.04$&$ 1.31\pm 0.02$& III\\
5576010116	 &59898.4921 &	779 	&0	&$0.79\pm0.06$ &$   3.11\pm 0.22$&$ 1.03\pm 0.14$& III\\
5576010117	 &59901.3300 &	1209 	&0	&$0.77\pm0.03$ &$   3.87\pm 0.25$&$ 0.93\pm 0.07$& III\\
5020560118	 &59902.4340 &	375 	&0	&$0.85\pm0.05$ &$   2.70\pm 0.42$&$ 1.01\pm 0.18$& III\\
5576010121	 &59913.4562 &	909 	&0	&$0.84\pm0.03$ &$   3.19\pm 0.27$&$ 0.62\pm 0.06$& III\\
5020560122	 &59924.0454 &	760 	&0	&$0.91\pm0.07$ &$   3.62\pm 0.26$&$ 0.93\pm 0.09$& III\\
5576010122	 &59925.2061 &	2078 	&0	&$0.80\pm0.03$ &$   2.95\pm 0.12$&$ 0.82\pm 0.06$& III\\
5576010123	 &59929.9123 &	7 	    &0	&$0.49\pm0.23$ &$   2.89\pm 0.57$&$ <0.69 $      & III\\
5576010124	 &59930.9444 &	37 	    &0	&$0.53\pm0.19$ &$   3.55\pm 0.46$&$ <0.57 $      & III\\
5576010125	 &59931.0090 &	36 	    &0	&$0.67\pm0.14$ &$   3.78\pm 0.56$&$ <0.25 $      & III\\
5576010126	 &59932.9440 &	42    	&0	&$0.74\pm0.21$ &$   2.69\pm 0.49$&$ <0.38 $      & III\\
5576010127	 &59933.0087 &	31 	    &0	&$0.61\pm0.16$ &$   2.50\pm 0.47$&$ <0.71 $      & III\\
5576010128	 &59934.2338 &	64 	    &0	&$0.81\pm0.17$ &$   2.67\pm 0.50$&$ <0.83 $      & III\\ 
\hline
\hline
\multicolumn{8}{l}{NOTES--}\\
\multicolumn{8}{l}{The errors are calculated with 1-$\sigma$ level uncertainties.}\\
\multicolumn{8}{l}{$^{\rm a}$ Obs. ID-NN, NN represents the serial number of the good time intervals (GTIs).}\\
\multicolumn{8}{l}{$^{\rm b}$ The start time of each observation or GTI.}\\
\multicolumn{8}{l}{$^{\rm c}$ The number of the bursts with a significance higher than 3-$\sigma$.}\\
\multicolumn{8}{l}{$^{\rm d}$ The 0.3--12 keV net count rate after removing bursts.}\\
\multicolumn{8}{l}{$^{\rm e}$ The 0.8--12 keV unabsorbed flux in units of $\rm 10^{-11} erg \, s^{-1} \, cm^{-2}$.}\\
\multicolumn{8}{l}{$^{\rm f}$ The three temporal parts are defined from the pulse profiles analysis in Section \ref{sec:Results-1}.}
\end{longtable*}

\section{Results} \label{sec:ANALYSIS AND RESULTS}
\subsection{Timing analysis}\label{sec:Results-1}

\begin{deluxetable}{lc}
\tablenum{2}
\tablecaption{Best fits spin parameters of SGR J1935+2154 from the 2022 October 12 to December 21 (excluding glitch time interval). 
The reference epoch is designated as the phase zero of the pulse profile.} \label{tab:tabletiming}
\tablewidth{0pt}
\tablehead{
\colhead{Parameter} & \colhead{Value}
}
\startdata
R.A. (J2000)&19:34:55.68\\ 
Decl. (J2000)&21:53:48.2 \\
Solar ephemeris&DE405\\
Start (MJD)     &59864\\  
Finish (MJD)    &59934\\
Reference epoch (MJD)&59865\\ 
Spin frequency $f$ (Hz)& $0.30752804(6)$\\
Spin-down rate$\dot f\ (\rm Hz\, s^{-1})$&$-4.82(3)\times10^{-12}$\\
Second derivative$\ddot f\ (\rm Hz\, s^{-2})$&$-4.9(7)\times10^{-21}$	\\
\hline
$\chi^2/\rm dof$&56/43\\
\enddata
\end{deluxetable}

\begin{figure*}
    \centering
    \subfigure[]{%
    \label{fig:2D}%
    \includegraphics[width=0.98\columnwidth]{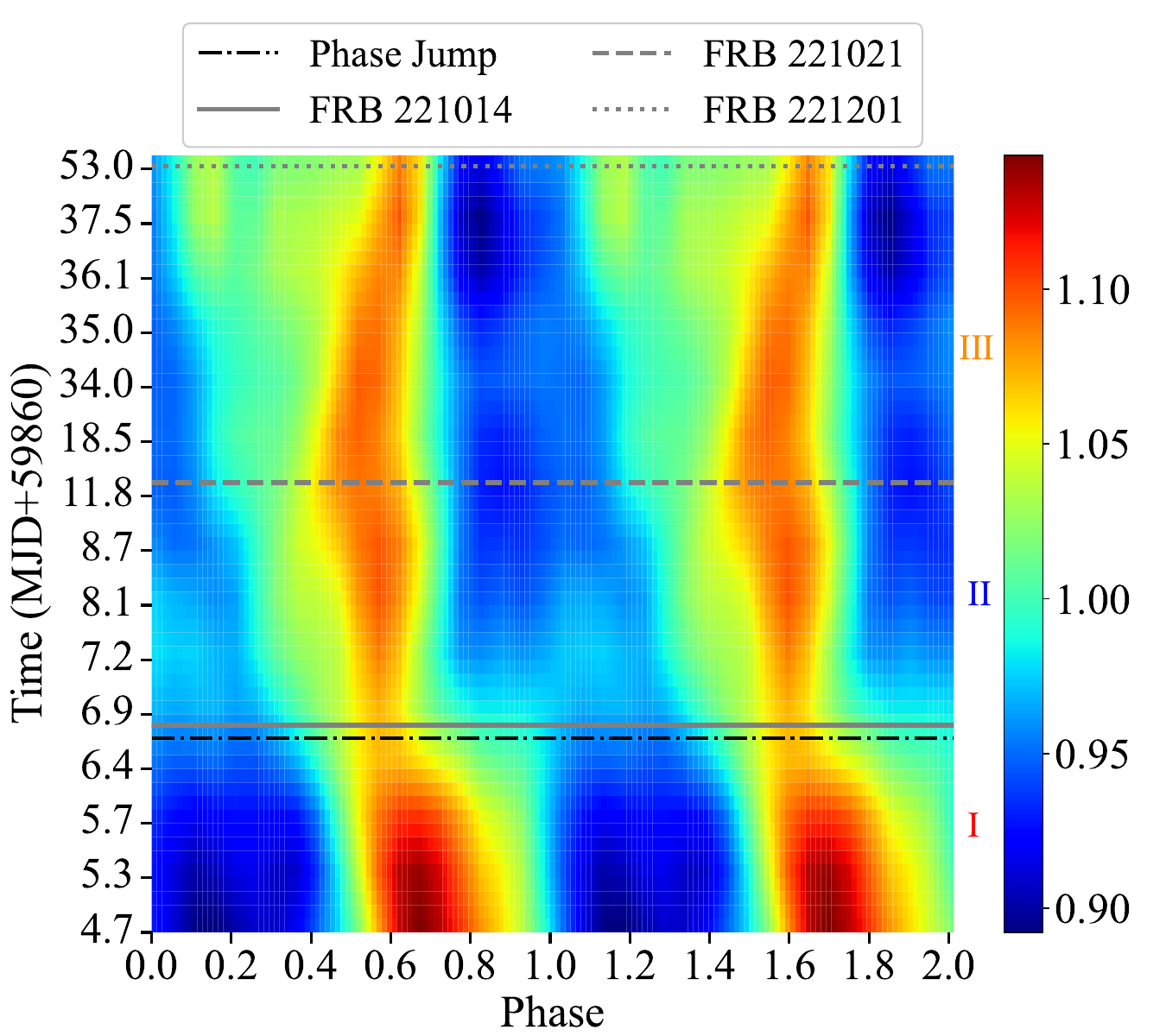}
  }  
      \subfigure[]{%
    \label{fig:profiles}%
    \includegraphics[width=0.98\columnwidth]{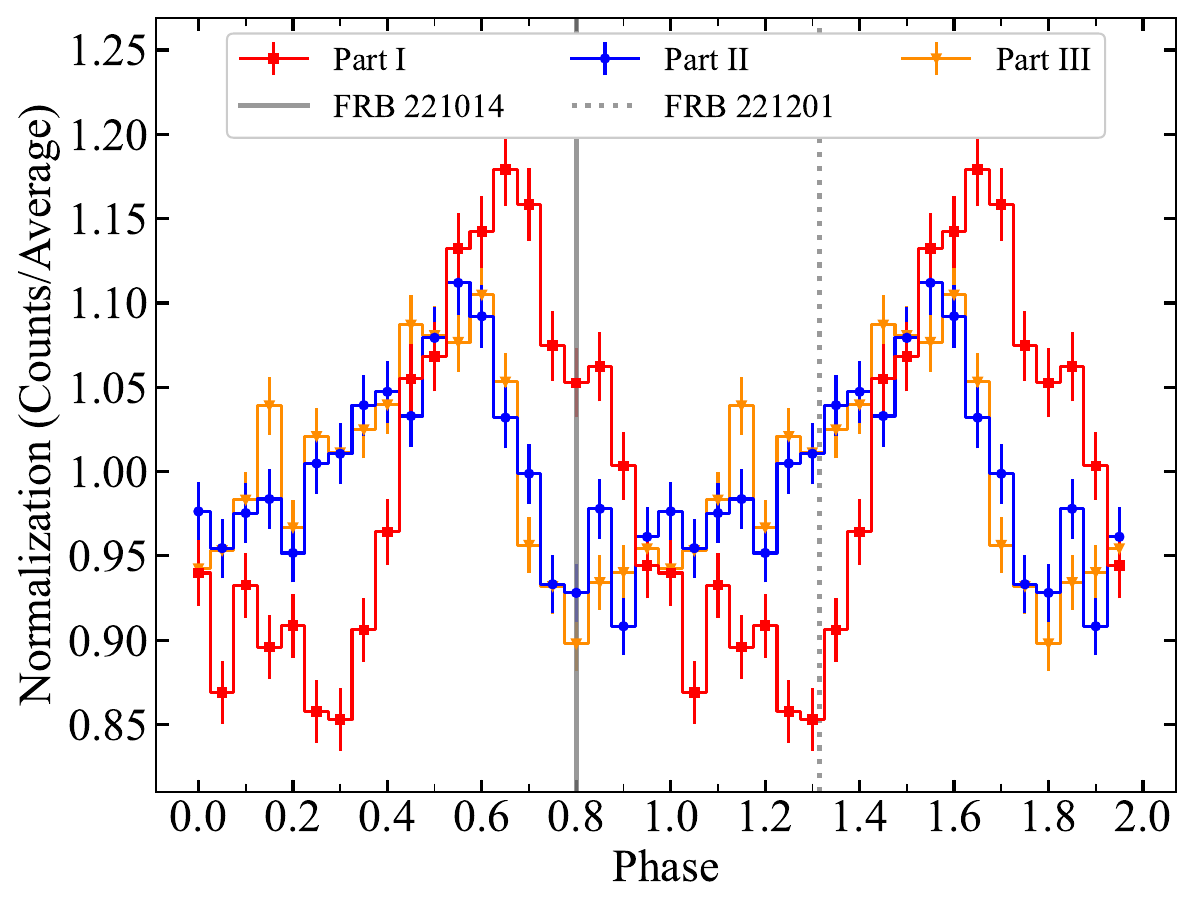}
  }  
\caption{The evolution of the pulse profiles. 
(a): The two-dimensional (2D) maps describe the evolution of the pulse profiles with time for NICER (0.8--4 keV).
The colors representing the values of the pulse profile are normalized by Pulse/Average count rate; the red represents pulse-on phase, and the blue represents pulse-off phase. 
20 bins within a phase are used to generate the pulse profiles, and the plot is smoothed through interpolation and Gaussian filtering for clarity. 
Phase zero is defined as the reference epoch MJD 59865 as shown in Table \ref{tab:tabletiming}.
The black dash-dotted line marks the position of the phase jump. 
The grey solid, dashed, and dotted lines mark the FRB 221014, FRB 221021, and FRB 221201, respectively.
Two cycles are shown for clarity.    
(b): The average pulse profiles of SGR J1935+2154. 
As shown in Table \ref{tab:table3}, Part I represents the pulse profile from the start to the phase jump. 
Part II represents the pulse profile from the phase jump to FRB 221021.
Part III represents the pulse profile from FRB 221021 to the end.
The errors are calculated with 1-$\sigma$ level uncertainties.} %
\label{fig:2D-profiles} %
\end{figure*}

\begin{figure*}
    \includegraphics[width=0.5\textwidth]{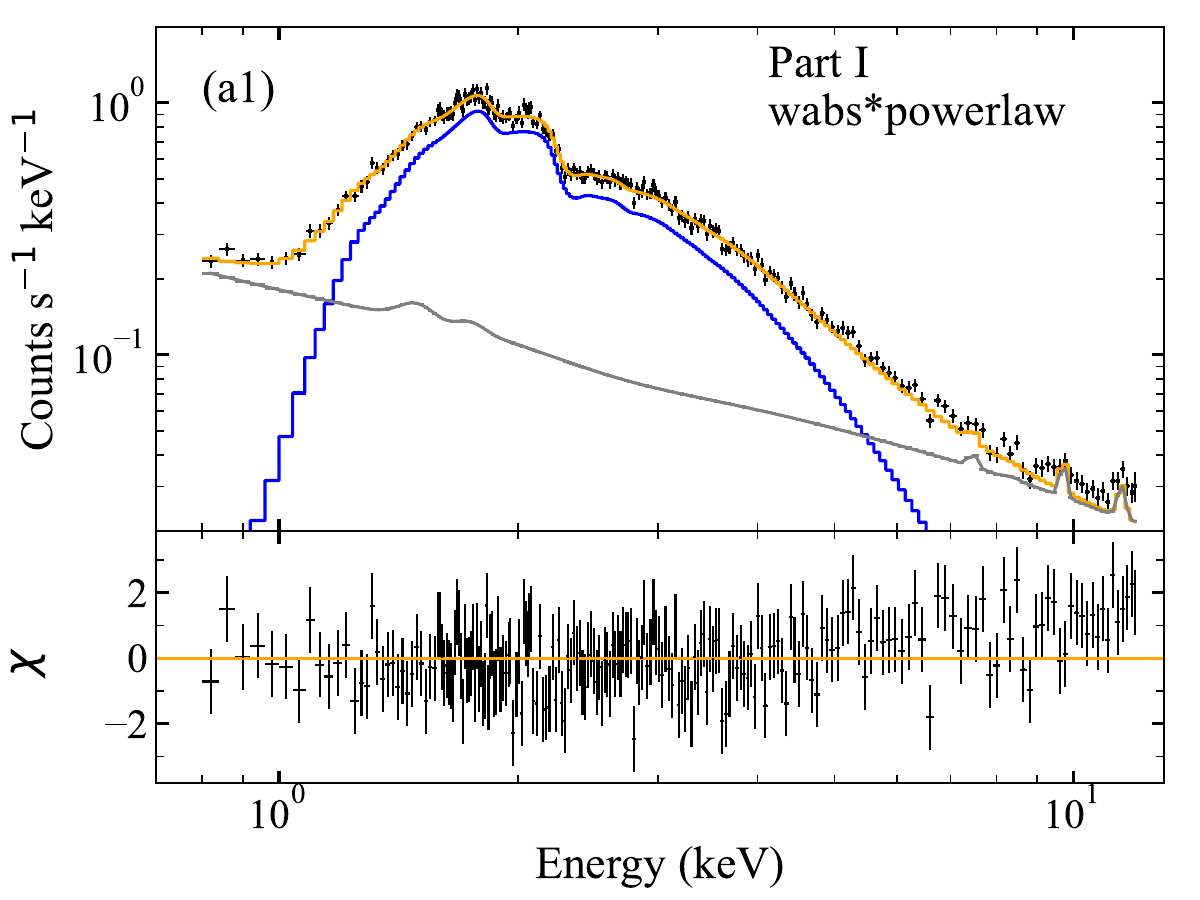}
	\includegraphics[width=0.5\textwidth]{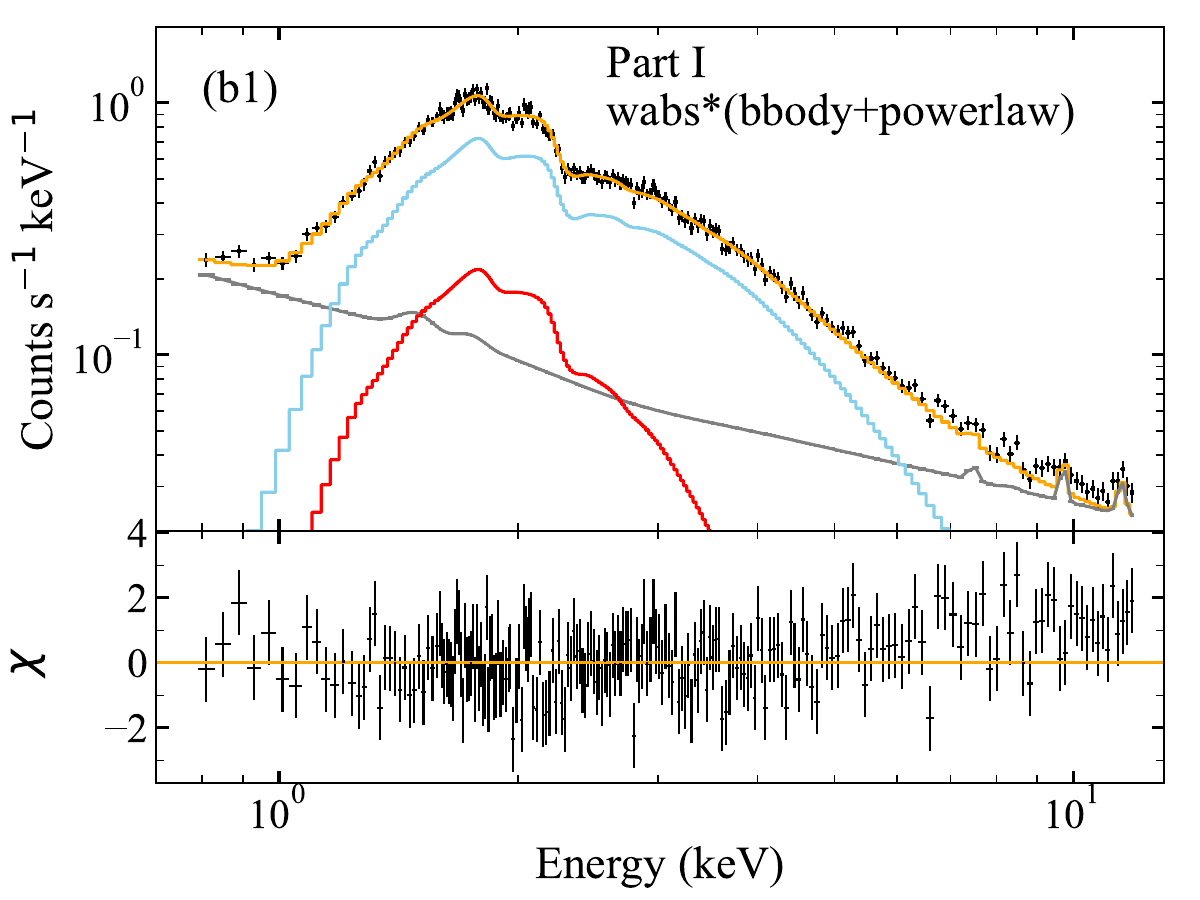}
	\qquad
	\includegraphics[width=0.5\textwidth]{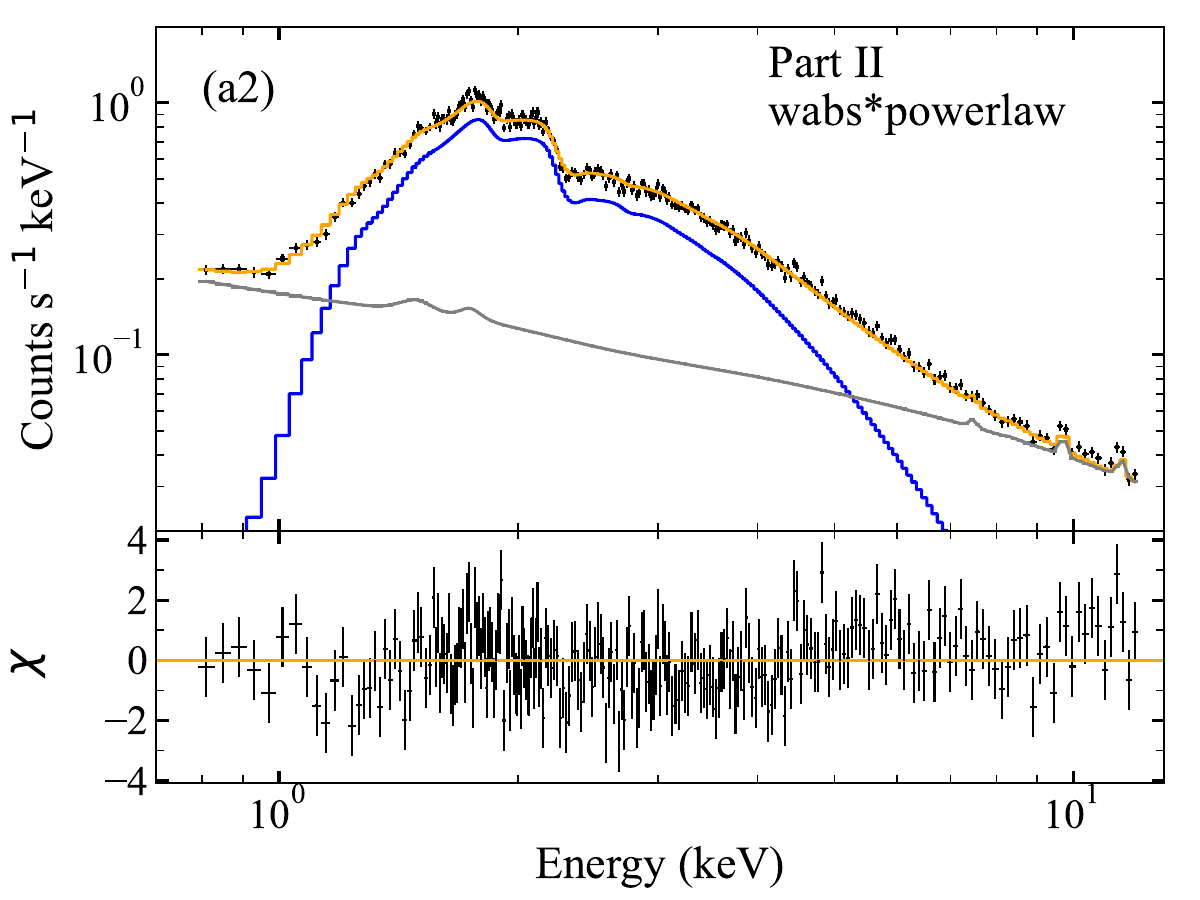}
	\includegraphics[width=0.5\textwidth]{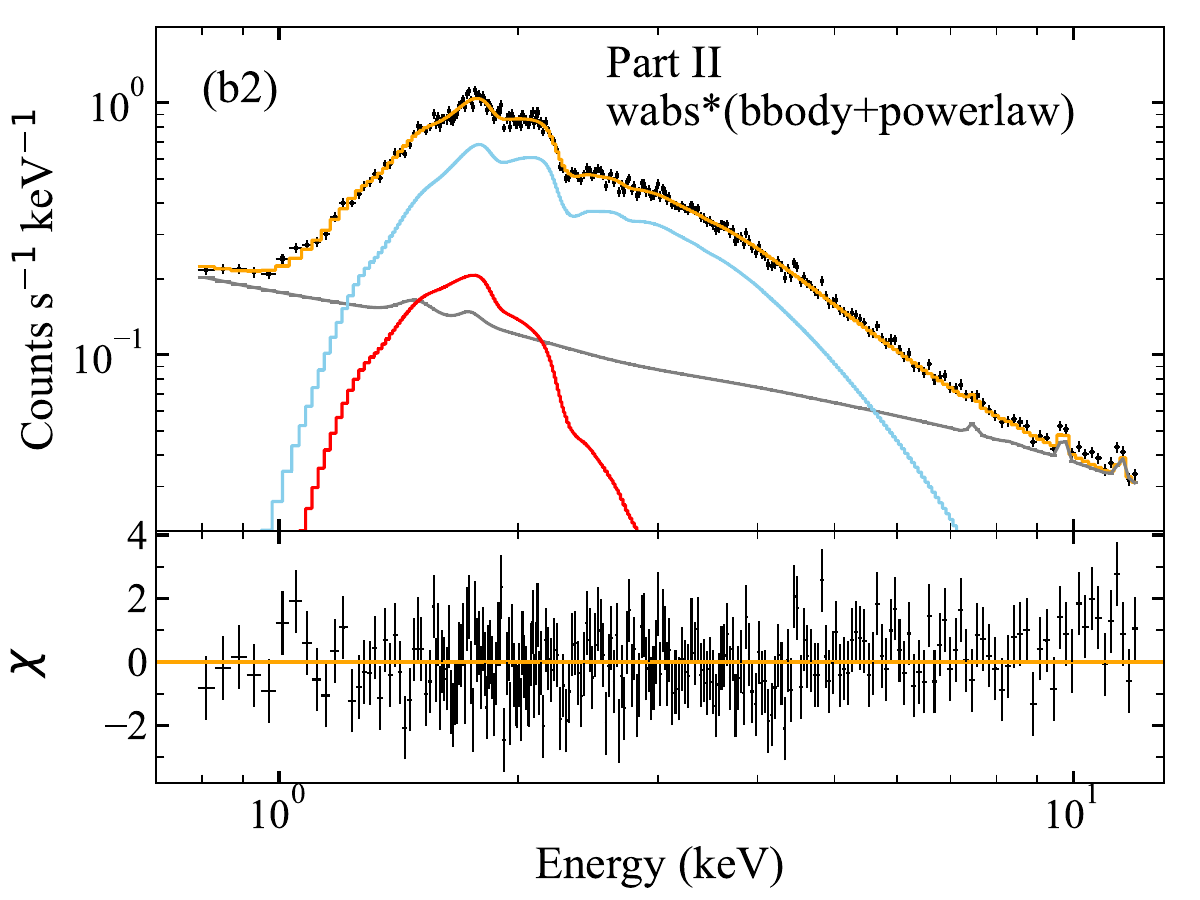}
	\qquad
	\includegraphics[width=0.5\textwidth]{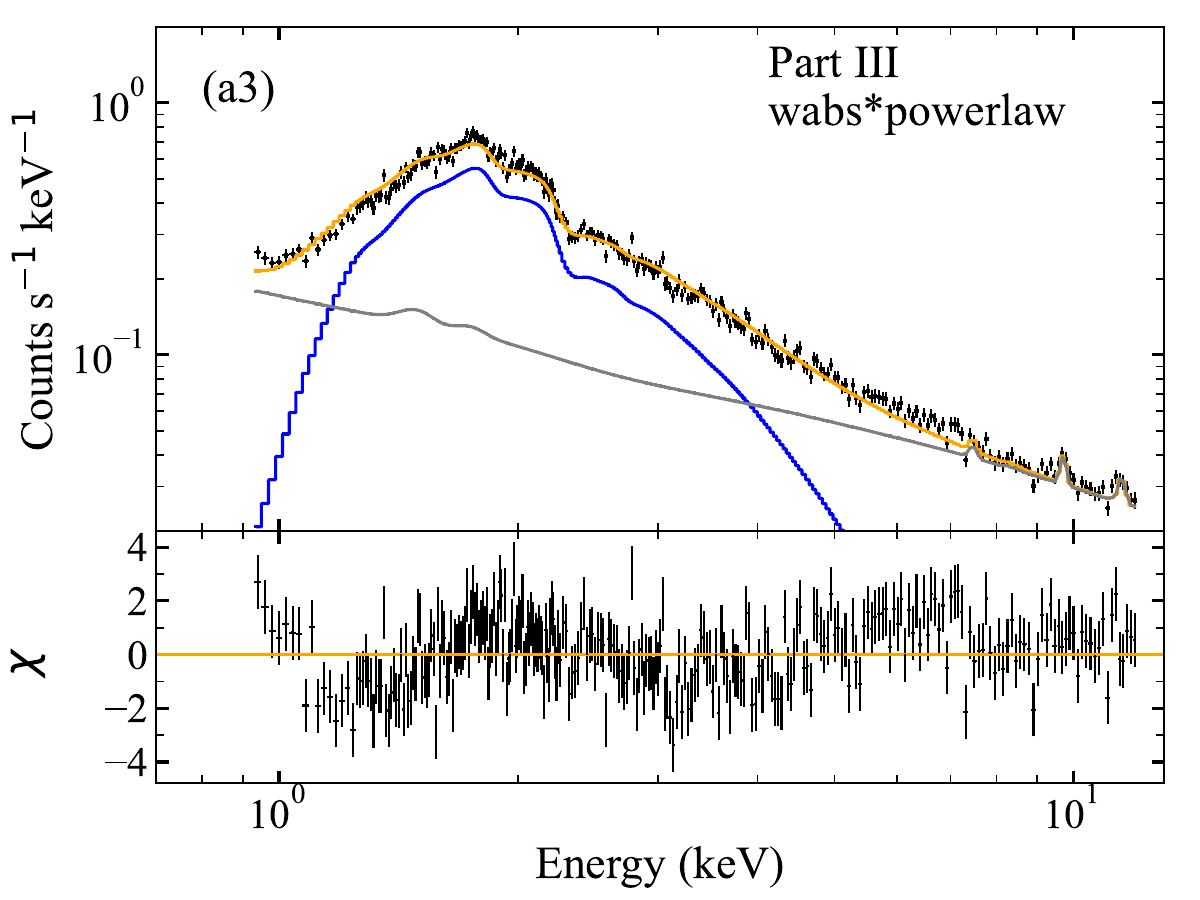}
	\includegraphics[width=0.5\textwidth]{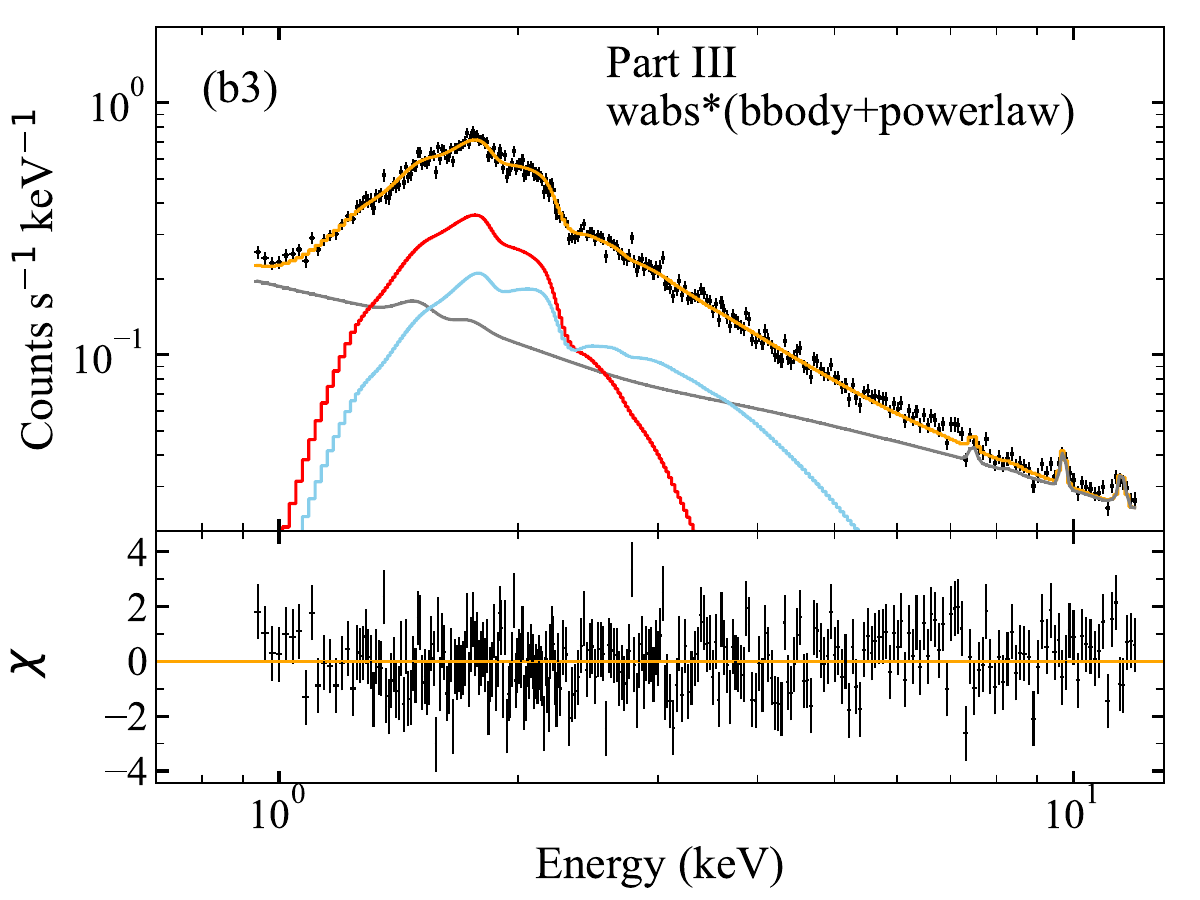}
    \caption{Spectra and residual of Part I (top figures), Part II (middle figures), and Part III (bottom figures). 
    Left figures: the total model (orange line) is plotted together with \texttt{powerlaw} (blue line) and the background model (grey line). Right figures: the total model (orange line) is plotted together with \texttt{powerlaw} (sky-blue line), \texttt{bbody} (red line), and the background model (grey line). The spectra are re-binned to display clarity.
    The errors are calculated with 1-$\sigma$ level uncertainties.}
    \label{fig:spec}
\end{figure*}

According to the study by \citet{2024Natur.626..500H}, there are double glitches around the high burst rate interval.
The timing parameter $\dot f$ between glitches jumps significantly, and the order of magnitude of $\dot f$ before the first and after the second glitches is consistent. 
Therefore, we deducted the data between the two glitches (MJD 59866.63--59866.99) to reduce their impact on the long-term evolution of the timing solution.
As shown in Table \ref{tab:tabletiming}, from the best-fitting of long-term data over a period of MJD 59864--59934, the spin frequency and the spin-down rate are calculated as $f=0.30752804(6)$ Hz and $\dot f=-4.82(3)\times10^{-12}$ Hz $\rm s^{-1}$, respectively.
This is slower than the spin measurement of $0.30789626(2)$ from the 2020 October outburst \citep{2023NatAs...7..339Y}, with a difference of 0.00036822(6) Hz.
The difference is consistent with a straightforward estimate; considering a two-year interval and only $\dot f$, the difference is calculated as $\sim$ 0.0003 Hz, which is consistent with the above values in terms of magnitude.

The pulse profiles exhibit a single-peaked shape in the 0.8--4 keV energy range, consistent with the result in \citet{2024RAA....24f5018L}.
As shown in Figure \ref{fig:2D}, the colors representing the values of the pulse profile are normalized by Pulse/Average count rate, the red represents pulse-on, and the blue represents pulse-off.
During the outburst, the shape of single-peaked profiles shows no significant change, the gradual broadening of the pulse in the green area in Figure \ref{fig:2D} is caused by a decrease in signal-to-noise ratio (SNR), and it cannot indicate that the profile has evolved.
However, the phase of the pulse profile changes significantly, the phase of the main peak remains at $\sim$ 0.65 before MJD 59866.62, while the phase of the main peak remains at $\sim$ 0.50 after MJD 59866.73.
Between the MJD 59866.62 and MJD 59866.73, a phase jump seems to occur, with a deviation of $\sim$ 0.15, which is consistent with Extended Data Figure 3 of \citet{2024Natur.626..500H}.
For the convenience of representation, we use the midpoint of this time interval (MJD 59866.68) as the time when the phase jump occurs marked with the black dash-dotted line.
The grey solid, dashed, and dotted lines mark the positions of the FRB 221014, FRB 221021, and FRB 221201, respectively.
The phase jump and FRB 221021 divide the data into three parts: Part I, Part II, and Part III.

As shown in Figure \ref{fig:profiles}, the average pulse profiles of there parts are displayed.
Part I represents the pulse profile from the start ($\sim$ MJD 59864.73) to the phase jump ($\sim$ MJD 59866.68),
Part II represents the pulse profile from the phase jump to FRB 221021 ($\sim$ MJD 59873.42),
and Part III represents the pulse profile from FRB 221021 to the end ($\sim$ MJD 59934.24).
The intensity of the pulse profiles are normalized by their average counts.
Phase zero is defined as the reference epoch of ephemeris as shown in Table \ref{tab:tabletiming}.
The pulse profiles of Part I and Part II are similar in shape, but there is a significant deviation in phase, and the main peak phase of Part II shifts $0.16\pm0.03$ to the left.
After ignoring statistical fluctuations, the pulse profiles of Part II and Part III are basically consistent.

The position of the radio bursts on the 0.8--4 keV X-ray pulse profile of persistent emission is also marked in Figure \ref{fig:profiles}.
The phase of radio bursts is calculated using the timing solution obtained from X-ray.
Consistent with the result from \citet{2024Natur.626..500H}, FRB 221014 appears at the position of pulse-off after the peak (part II), which seems to be consistent with the radio burst B1 in Figure 5 by \citet{2020ApJ...904L..21Y}.
FRB 221201 appears in the rising phase before the peak (part III), which seems to be consistent with the radio burst B2 in Figure 5 by \citet{2020ApJ...904L..21Y}.
It should be noted that the position of FRB 221201 is greatly influenced by the ephemeris $\dot f$ or $\ddot f$, as it spans a large time interval from the reference epoch.

\subsection{Spectral analysis}\label{sec:Results-2}
\subsubsection{Phase-averaged spectral analysis}\label{sec:spec-2-1}

\begin{deluxetable*}{lcccccc}
\tablenum{3}
\tablecaption{Results of the spectral parameters presented in Section \ref{sec:Results-2}.  \label{tab:spec}}
\tablehead{
\colhead{Model}& \multicolumn{3}{c}{Model I, \texttt{wabs}*\texttt{powerlaw}}&\multicolumn{3}{c}{Model II, \texttt{wabs}*(\texttt{bbody}+\texttt{powerlaw})}\\
\colhead{Epoch}& \colhead{Part I}& \colhead{Part II} & \colhead{Part III}& \colhead{Part I} & \colhead{Part II}&\colhead{Part III}}
\startdata
$kT_{\rm BB}$ (keV)        &-&-&-&$0.43\pm0.02$ & $0.37\pm0.02$ & $0.35\pm0.01$\\ 
R$\rm ^{ a} _{BB}$ (km)    &-&-&-&$1.9\pm0.1$   &$2.9\pm0.2$    &$4.2\pm0.2$  \\
$\Gamma$&$2.30\pm0.01$&$2.09\pm0.01$&$2.95\pm0.02$&$2.08\pm0.06$ &$1.81\pm0.07$  &$2.00\pm0.14$   \\
Norm$^{\rm b}$ PL          &$6.7\pm0.1$&$6.1\pm0.1$&$6.9\pm0.1$&$4.5\pm0.4$   &$3.7\pm0.3$    &$1.6\pm0.2$    \\
Total Flux$^{\rm c}$       &$2.12\pm0.01$&$2.26\pm0.01$&$1.32\pm0.01$&$2.06\pm0.01$ &$2.23\pm0.01$  &$1.19\pm0.02$  \\
BB Flux$^{\rm c}$          &-&-&-&$0.22\pm0.02$ &$0.28\pm0.01$  &$0.51\pm0.01$  \\
PL Flux$^{\rm c}$          &-&-&-&$1.85\pm0.02$ &$1.94\pm0.02$  &$0.68\pm0.02$  \\
Ratio$^{\rm d}$ ($\%$)     &-&-&-&$10.7\pm0.9$  &$12.6\pm0.5$   &$42.9\pm1.1$   \\
$\chi^2/\rm dof$           &1596/1418&1395/1418&1270/1099&1591/1416     &1363/1416      &1127/1097      \\
\enddata
\tablecomments{\\
The errors are calculated with 1-$\sigma$ level uncertainties.\\
$^{\rm a}$ The source radius in km at the distance of 6.6 kpc \citep{2020ApJ...905...99Z}.\\
$^{\rm b}$ The normalization of the power law in units of $\times 10^{-3} \rm photons \cdot keV^{-1}\,cm^{-2}\,s^{-1}$.\\
$^{\rm c}$ The 0.8--12 keV unabsorbed total, BB and PL flux in units of $\times \rm 10^{-11} erg \, s^{-1} \, cm^{-2}$.\\
$^{\rm d}$ The ratio of the \texttt{bbody} flux to total flux.}

\end{deluxetable*}

The selection of spectral models has been systematically studied during the previous outbursts; the spectra below 10 keV are described by the combination of \texttt{bbody} and \texttt{powerlaw} \citep{2017ApJ...847...85Y, 2020ApJ...904L..21Y, 2022MNRAS.516..602B}.
\citet{2024arXiv241000635S} also reported that most of the spectra (in the range of 1--5 keV) could be better described with an absorbed \texttt{bbody} plus \texttt{powerlaw} model during 2022 outburst.
We attempted to fit the spectra in 0.8--12 keV energy range using different model combinations of \texttt{bbody} and \texttt{powerlaw}.
Based on the selection criteria of residual showing no significant structure, combined with $\chi^2/\rm dof$ values less than 1.3 and the null hypothesis probability for the improvement from F-test less than 0.05, we determined that \texttt{wabs}*\texttt{powerlaw} (Abbreviated as Model I) and \texttt{wabs}*(\texttt{bbody}+\texttt{powerlaw}) (Model II) were the best choices for fitting spectra.
The hydrogen column density of \texttt{wabs} is fixed at $N_{\rm H}=2.3\times10^{22}\,\rm cm^{-2}$ according to the previous studies \citep[e.g.][]{2017ApJ...847...85Y, 2018MNRAS.474..961C, 2022MNRAS.516..602B}.

In Figure \ref{fig:spec}, the average spectra of Part I, Part II, and Part III fitted with these two model combinations are displayed.
The reason why the residuals are higher on the high energy side is probably because the hard power law above 10 keV is contaminating the data.
The Chi-squared test indicates that there is no significant difference between the two combinations for Part I and Part II.

For Part I, the $\chi^2/\rm dof$ of Model I and Model II are $1596/1418\approx1.126$ and $1591/1416\approx1.124$, respectively, as shown in Figure \ref{fig:spec} (a1) and (b1).
The Chi-squared statistics of Model I and Model II are almost equivalent, thus making it impossible to distinguish which model is the best fit based on this criterion.
The F-test gives a null hypothesis probability for the improvement of the fit of 0.11. 
Considering the inclusion of \texttt{bbody}, it is not statistically necessary, but is needed in the fit. 
This is because \texttt{bbody} has been observed in previous outbursts \citep{2017ApJ...847...85Y, 2020ApJ...904L..21Y, 2022MNRAS.516..602B} and at other temporal parts within this outburst, indicating its possible physical presence.
However, due to spectral variability, its statistical significance may not be prominent.
The flux of \texttt{bbody} (red line) is one order of magnitude smaller than the flux of \texttt{powerlaw} (sky-blue line), thus \texttt{bbody} does not have a significant impact on the residual of spectra 

For Part II, the $\chi^2/\rm dof$ of Model I and Model II are $1395/1418\approx0.983$ and $1363/1416\approx0.960$, respectively, as shown in Figure \ref{fig:spec} (a2) and (b2). 
The F-test gives a null hypothesis probability for the improvement of the fit of $7.32\times10^{-8}$, indicating that the inclusion of the \texttt{bbody} component is statistically significant and thus necessary in the model.

For Part III, the $\chi^2/\rm dof$ of Model I and Model II are $1270/1099\approx1.16$ and $1127/1097\approx1.03$, respectively.
The obvious residual structure in Figure \ref{fig:spec} (a3) and (b3) demonstrates the significance of the \texttt{bbody} component.
Additionally, the flux of \texttt{bbody} accounts for $42.9\pm1.1\%$ of the total flux.

As shown in Table \ref{tab:spec}, the decrease in the total flux of Part III relative to Part I and Part II is mainly contributed by the decrease in \texttt{powerlaw}, while accompanied by a slight increase in \texttt{bbody}.
The temperature of \texttt{bbody} slightly decreases, while its radius grows from $\sim 2$ km to $\sim 4$ km.
The proportion of \texttt{bbody} component increases from $\sim 10\%$ to $\sim 40\%$ (\texttt{powerlaw} component proportion decreases from $\sim 90\%$ to $\sim 60\%$), which is consistent with the results of 2020 outburst by \citet{2022MNRAS.516..602B}, where the \texttt{powerlaw} component proportion decreased from $\sim 75\%$ to $\sim 45\%$.

\subsubsection{Spectral evolution}\label{sec:spec-2-3}

\begin{figure*}
    \includegraphics[width=\textwidth]{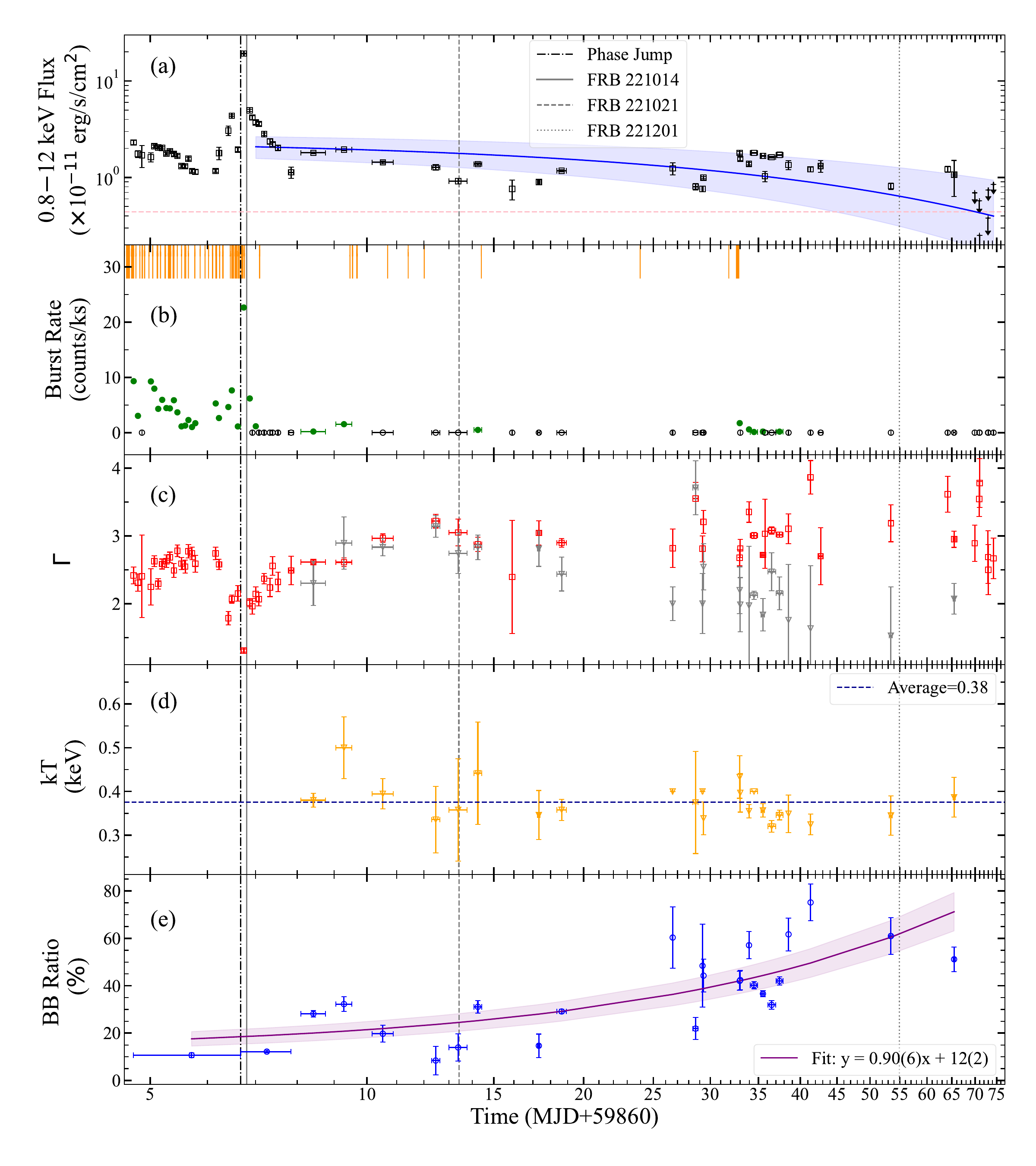}
    \caption{Evolution of the flux, burst rate, photon index ($\Gamma$), and the ratio of the \texttt{bbody} flux to the total flux with time.
    Panel (a): the evolution of the unabsorbed flux in units of $\rm 10^{-11} erg \, s^{-1} \, cm^{-2}$. The fluxes are computed in the energy range 0.8--12 keV.
    The blue line fits the exponential decay of the flux, with the result of $y= 2.5(2)e^{-0.025(5)x}$. 
    The pink dashed line represents the quiescent flux level as reported by \citet{2024arXiv241000635S}.
    Panel (b): the evolution of the burst rate. Bursts are identified with a significance higher than 3-$\sigma$.
    Black hollow points represent zero.
    The orange arrows represent the identified bursts from Fermi/GBM.
    Panel (c): the evolution of the photon index of power law.
    The red data points come from Model I (PL), and the gray points come from Model II (BB+PL).
    Panel (d): the evolution of the temperature of BB.
    Panel (e): the evolution of the ratio of the \texttt{bbody} flux to the total flux.
    The errors are calculated with 1-$\sigma$ level uncertainties. 
    The black dash-dotted line marks the phase jump time of the pulse profile which is $\sim$ MJD 59866.68, and grey solid, dashed, and dotted lines mark the FRB 221014, FRB 221021, and FRB 221201, respectively.}
    \label{fig:par}
\end{figure*}

We perform the spectral fitting of each observation, and the results are shown in Table \ref{tab:table3}.
In Figure \ref{fig:par}, the evolution of the flux, burst rate, photon index ($\Gamma$), and the ratio of the \texttt{bbody} flux to the total flux with time are displayed from top to bottom panels.

In Figure \ref{fig:par} (a), the unabsorbed total flux initially remains at $\sim (1-2) \times \rm 10^{-11} erg \, s^{-1} \, cm^{-2}$.
Then there is a flare around the phase jump which is marked with the black dash-dotted line, and the flux increases more than ten times at the peak of the flare, reaching $\sim 19 \times \rm 10^{-11} erg \, s^{-1} \, cm^{-2}$. 
The phase jump is close to the peak of the flare, and it can be discerned that it precedes the peak.
The flare starts on $\sim$ MJD 59866.4 and ends on $\sim$ MJD 59867.2, lasting for less than a day.
The blue line fits the exponential decay of the flux, with the fitting result being $Flux (10^{-11})= 2.5(2)e^{-0.025(5)T(\rm days)}$, based on data after MJD 59867.068.
The characteristic time of the decay is $\sim$ 40 days.
The pink dashed line represents the quiescent flux level as reported by \citet{2024arXiv241000635S}.
It is worth noting that FRB 221014 \citep{2022ATel15697....1M, 2022ATel15681....1D} appears slightly behind the peak of the flare, marked with the grey solid line.
The grey dashed and grey dotted lines also mark the FRB 221021 and FRB 221201, respectively.

\begin{figure*}
    \includegraphics[width=\textwidth]{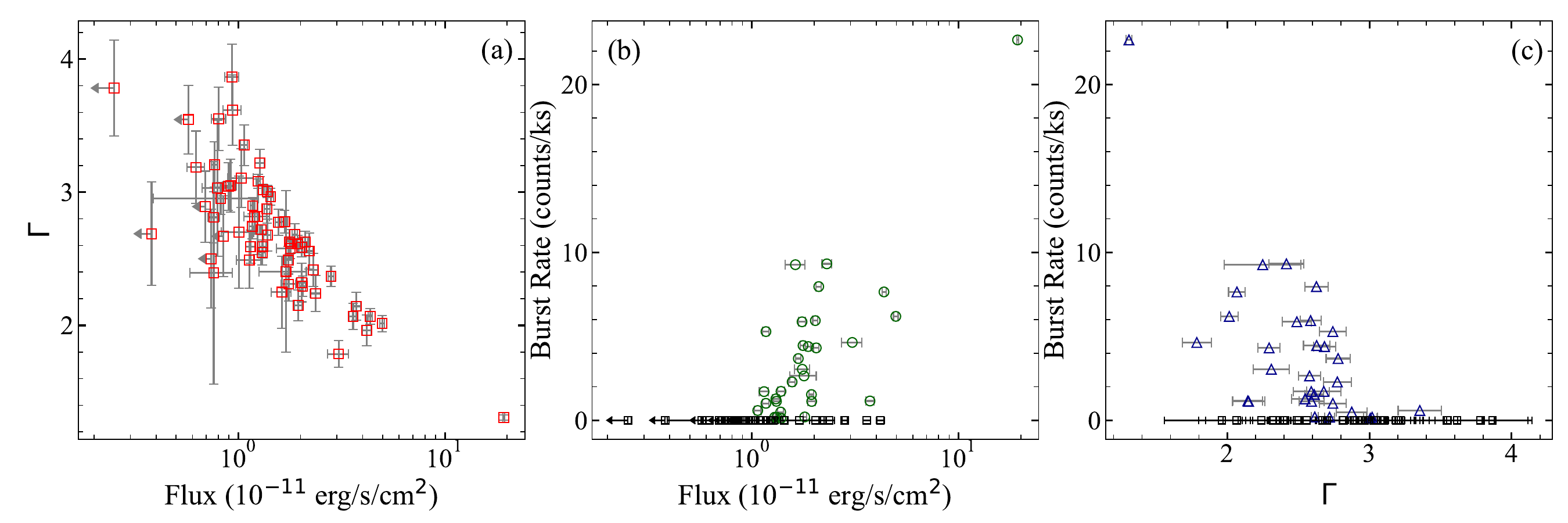}
    \caption{The correlation between the 0.8--12 keV flux, the photon index $\Gamma$, and the  burst rate in Figure \ref{fig:par}. In Panels (b) and (c), the black hollow square points represent the burst rate of zero. The errors are calculated with 1-$\sigma$ level uncertainties.}
    \label{fig:correlation}
\end{figure*}

In Figure \ref{fig:par} (b), the burst rate is defined as the number of bursts divided by the exposure time per observation or GTI.
From the start of the outburst, the burst rate decreases until the beginning of the flare, during which the burst rate significantly increases, and there is a high burst rate interval at the peak of the flare.
Higher energy bursts from Fermi/GBM are marked with orange arrows.
The orange arrows gather more densely just behind the black dash-dotted line, which also indicates that there is a high burst rate interval.
After the flare, the burst rate rapidly decreases to zero, and there are also almost no higher energy bursts.
There is a positive correlation between the burst rate and the flux as shown in Figure \ref{fig:correlation} (b).
The P-value of the Pearson correlation test is $2\times10^{-14}$, indicating a statistically significant positive correlation, with a correlation coefficient of 0.79.

In Figure \ref{fig:par} (c), the evolution of the photon index from Model I and II is displayed.
The red data points come from Model I (PL).
When the outburst decays normally, the photon index is $\sim$ 2.5 and slowly increases, but decreases to $\sim$ 1.3 during the flare.
After the flare, the photon index gradually increases and remains at $\sim$ 3.
The gray points come from Model II (BB+PL). 
When the proportion of BB is small ($\lesssim30\%$), it is consistent with the results of Model I.
And when the proportion of BB increases ($\gtrsim30\%$), the photon index remains at $\sim$ 2, which is smaller than the results of Model I.
After MJD 59875, the decrease in data SNR leads to the dispersion of the photon index and larger errors.
As shown in Figure \ref{fig:correlation} (a) and (c), the evolution of the photon index (Model II) is inversely proportional to the flux and the burst rate.
The Pearson correlation test yields a P-value of $8\times10^{-8}$ with a correlation coefficient of $-$0.61 for the flux, and a P-value of $4\times10^{-6}$ with a correlation coefficient of $-$0.54 for the burst rate, respectively.

In Figure \ref{fig:par} (d), the temperature of \texttt{bbody} varies between $\sim$ 0.3 keV and $\sim$ 0.5 keV, which is roughly consistent with the results from \citet{2024arXiv241000635S}.
There is no significant evolution in temperature, with an average value of 0.38 keV.

In Figure \ref{fig:par} (e), the ratio of the \texttt{bbody} flux to the total flux increases from $\sim10\%$ at the beginning to $\sim60\%$ at the end, with a maximum value of $75\pm7\%$.
A linear model is used to fit this growth trend, and the fitting result is $Ratio(\%)=0.90(6)T(\rm days)+12(2)$,
which means that the \texttt{bbody} ratio is growing at a rate of $\sim1\%$ per day.
Some missing data points are due to the poor data quality of these observations which is not sufficient to limit the complexity of Model II.

\begin{figure*}
    \centering
    \includegraphics[width=\textwidth]{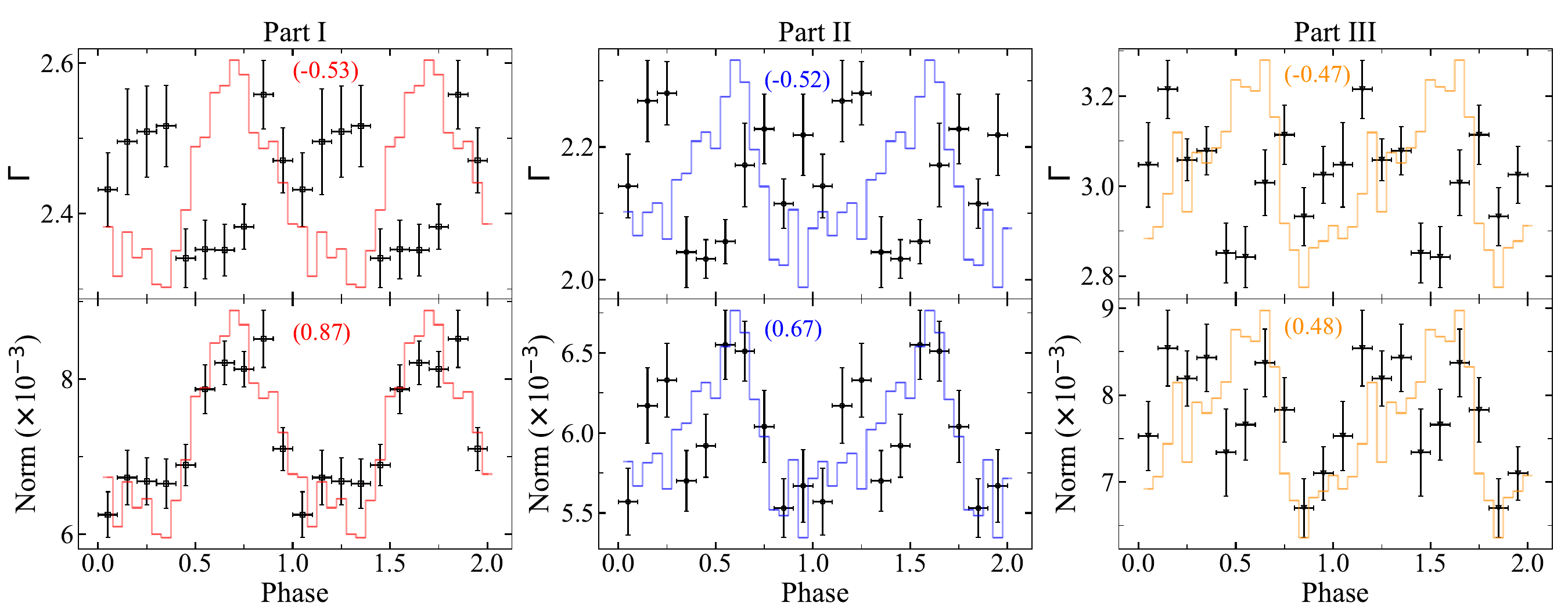}
    \caption{Fitting results of the phase-resolved spectral analysis of Model I.
    Solid lines represent the pulse profiles as shown in Figure \ref{fig:profiles}.
    The errors are calculated with 1-$\sigma$ level uncertainties. The values in parentheses represent the correlation coefficient between the pulse and parameters, with positive numbers indicating positive correlation and negative numbers indicating negative correlation.}
    \label{fig:phase-resolved}
\end{figure*}

\begin{figure*}
    \centering
    \includegraphics[width=\textwidth]{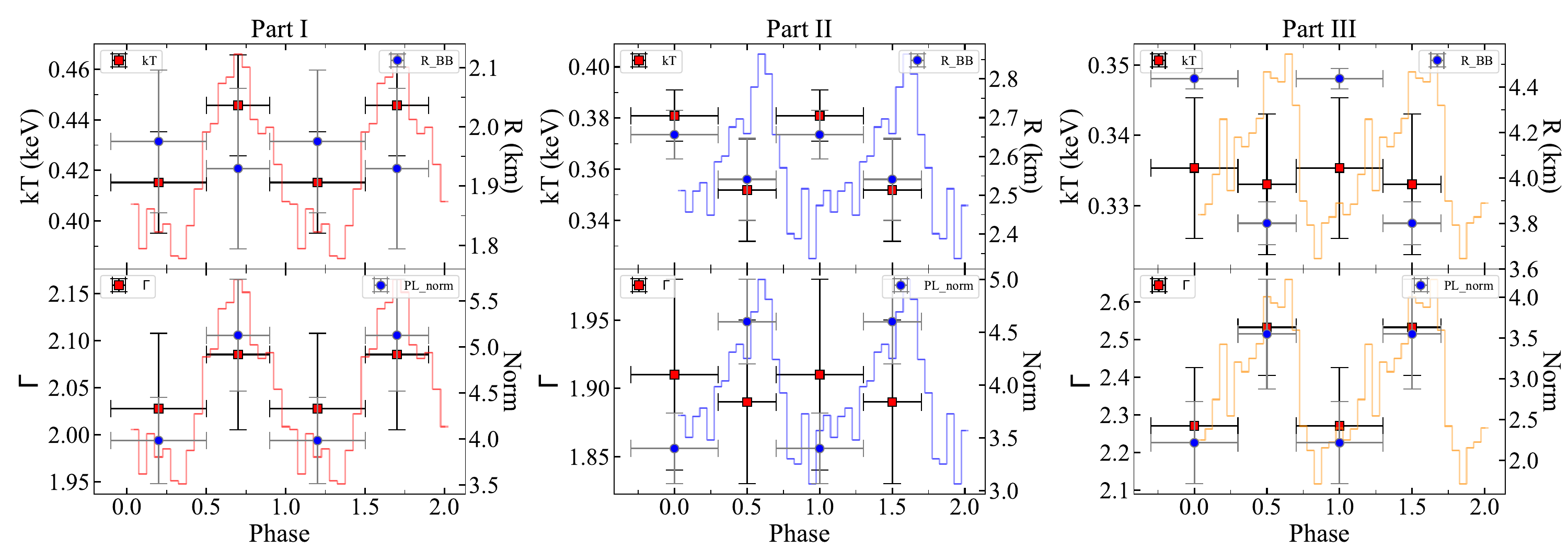}
    \caption{Fitting results of the phase-resolved spectral analysis of Model II.
    Solid lines represent the pulse profiles as shown in Figure \ref{fig:profiles}.
    The errors are calculated with 1-$\sigma$ level uncertainties.
    The source radius in km at the distance of 6.6 kpc \citep{2020ApJ...905...99Z}.
    The normalization of the power law in units of $10^{-3} \rm photons \cdot keV^{-1}\,cm^{-2}\,s^{-1}$.}
    \label{fig:phase-resolved-BB}
\end{figure*}

\subsubsection{Phase-resolved spectral analysis}\label{sec:spec-2-2}
In order to analyze the possible transition of the spectral property before and after the phase jump, we study the phase-resolved spectral analysis on the average spectra of Part I, Part II, and Part III, respectively.

For Model I, the entire cycle is divided into ten phase intervals.
In Figure \ref{fig:phase-resolved}, the photon index and the normalization of the power law of three parts are presented, respectively.
For Part I, there is an inverse correlation between the photon index and the pulse profile, with a correlation coefficient of $-$0.53.
The photon index varies between 2.34 and 2.55, with the minimum pulse ($\sim 0.3$) corresponding to $2.52\pm0.06$ and the maximum pulse ($\sim 0.7$) corresponding to $2.35\pm0.04$. 
The normalization is proportional to the pulse profile, with a correlation coefficient of 0.87.
For Part II, there is also an inverse correlation between the photon index and the pulse profile, with a correlation coefficient of $-$0.52.
The photon index varies between 2.03 and 2.28, with the minimum pulse ($\sim 0.9$) corresponding to $2.22\pm0.05$ and the maximum pulse ($\sim 0.6$) corresponding to $2.05\pm0.03$.
The normalization is proportional to the pulse profile, with a correlation coefficient of 0.67.
For Part III, the evolution is similar to Part II, and the correlation coefficient between the photon index and the pulse is $-$0.47.
The normalization is proportional to the pulse profile, with a correlation coefficient of 0.48.

In brief, when the phase jump occurs (Part I and Part II), the spectral parameters also correspondingly change, when the pulse phase remains constant (Part II and Part III), the spectral parameters also approximately remain constant.
The evolution of spectral parameters with phase does not change.

For Model II, the entire cycle is divided into two phase segments, pulse on and pulse off. 
As shown in Figure \ref{fig:phase-resolved-BB}, there is no significant change in $kT_{\rm BB}$ and $\Gamma$ between pulse on and pulse off for all three parts, which is consistent with the result from \citet{2024ApJ...965...87I}. 
The source radius $\rm R_{BB}$ remains consistent across the different phase segments for Parts I and II, but for Part III, the $\rm R_{BB}$ of pulse off is larger than that of pulse on, increasing by $\sim$15\%.
Meanwhile, for three Parts, from pulse on to pulse off, the PL normalization is decreasing.

\section{Discussion} \label{sec:Discussion}
We analyzed the NICER observations of SGR J1935+2154 during its 2022 active period and performed the timing and spectral analysis of the X-ray persistent emission.
During the decay of the outburst, we found a phenomenon of the phase jump of the pulse profile, with a deviation of $0.16\pm0.03$, as shown in Figure \ref{fig:2D-profiles}.
Before and after the phase jump, the phase-resolved spectral properties also changed correspondingly, as shown in Figure \ref{fig:phase-resolved}. 
Around the phase jump, a flare appeared on the light curve accompanied by changes in spectral properties.
The phase jump is slightly earlier than the flare, and FRB 221014 is closely following the flare, as shown in Figure \ref{fig:par}.
Furthermore, there are no significant changes in the properties of X-ray persistent emission near FRB 221021 and FRB 221201.

The increase of the persistent X-ray flux is quite common in magnetars, accompanied by spectral hardening, manifested as a decrease in the power law (PL) index or an increase in the blackbody (BB) temperature \citep[e.g.,][]{2010MNRAS.408.1387I, 2018MNRAS.474..961C}.
This spectral evolution is also ubiquitous in the previous active period of SGR J1935+2154 \citep{2016MNRAS.457.3448I, 2017ApJ...847...85Y, 2020ApJ...904L..21Y}, as well as other sources such as 1E 1547--5408 \citep{2011ApJ...739...94S} and 1E 2259+586 \citep{2003ApJ...588L..93K}.
In this study, the spectra are well modeled by a single PL with a photon index $\Gamma \sim (1.3-3.0)$. 
The Chi-squared statistics and the insensitive fitting of the BB indicate that the thermal component is not important.
These characteristics are slightly different from the 2020 outburst of SGR J1935+2154, where a thermal component is required to model the spectra \citep{2020ApJ...904L..21Y, 2022MNRAS.516..602B}.
In the early observation of the 2020 outburst, the proportion of PL accounting for $\sim 75\%$ \citep{2022MNRAS.516..602B} is less than $\sim 90\%$ of the 2022 outburst, which could explain the insensitivity of the BB component in the spectral fitting in this study.
In Figure \ref{fig:par}, the PL photon index fitting results of Model I and Model II show that when the proportion of BB is greater than $\sim 30\%$, the thermal component will affect the fitting of the photon index. When the proportion of BB is less than $\sim 30\%$, the influence of thermal components on the photon index is negligible.
The proportion of BB seems to have a potential relationship with the burst rate. 
On the $\sim$ MJD 59895, where the burst rate increases, the BB ratio decreases, with a slight increase in flux, indicating the possible existence of a small flare here.

The combination of a BB and a PL component is generally interpreted as the thermal emission from the surface of NS, and the scattering of seed photons by physical mechanisms taking place in the magnetosphere, such as the Resonant Cyclotron Scattering \citep{2008MNRAS.386.1527N} and the currents scattering in the twisted magnetosphere model \citep{2002ApJ...574..332T}.
The degree of scattering increases with the increasing magnetic twist, which could be localized to a restricted portion of the magnetosphere \citep[e.g.,][]{2007ApJ...659L.141O, 2009ApJ...703.1044B}.
The persistent emission during the SGR J1935+2154 outburst dominated by the PL component ($\sim 90\%$) indicates that the twisted magnetic field plays an important role in the outburst.
During the decay of the outburst, the temperature and flux of BB do not show a significant evolution, while the flux of PL shows a decreasing trend.
This similar phenomenon also has been observed in 2014, 2015, and 2016 outbursts; the flux of BB remains at $\sim 1.5\times \rm 10^{-12} erg \, s^{-1} \, cm^{-2}$, and the flux of PL decreases from  $\sim (3-8)\times \rm 10^{-12} erg \, s^{-1} \, cm^{-2}$ to $\sim 1\times \rm 10^{-12} erg \, s^{-1} \, cm^{-2}$ \citep{2017ApJ...847...85Y}.
According to the spectral results obtained by \citet{2024Natur.626..500H}, during the early stage of outburst decay, the BB temperature first increases and then decreases with the flare, with this rapid decline terminating approximately 10 hours after the FRB. 
This rapid decrease is consistent with the result of the 2020 outburst, where BB temperature rapidly decreased within 3 days after the outburst, then entered a slow decrease stage lasting for several tens of days \citep{2020ApJ...904L..21Y}.

The pulse profile exhibits a single-peaked shape without a significant variability in the 0.8--4 keV energy band, consistent with the previous observations \citep{2022MNRAS.516..602B, 2023NatAs...7..339Y}. 
However, particularly, a phase jump was discovered around the peak of a flare during this outburst.
Continuous observations indicate that the phase jump occurs between MJD 59866.62 and MJD 59866.73, which coincides with the first glitch time (MJD 59866.63) from NICER and NuSTAR data \citep{2024Natur.626..500H}. 
Figure \ref{fig:profiles} shows that the overall left shift of the pulse profile is 0.16(3) phase, including the pulse-on and the pulse-off.
The phase-resolved spectral analysis also supports the phase jump, before and after the phase jump, the fitting parameters of phase-resolved spectra always maintain a constant correlation with the pulse profile. 
In other words, the spectral properties also shift 0.16(3) phases to the left after the phase jump.
The consistency between phase jump and glitch time indicates that this is due to the glitch in the timing solution, which caused by magnetospheric wind \citep{2023NatAs...7..339Y, 2024Natur.626..500H}.
However, at the second glitch time (MJD 59866.99), there is no obvious phase jump here, which may be due to data quality blurring possible phase jumps, but it may also indicate that only a few special glitches are associated with the phase jumps, indicating that there may be other physical mechanisms that cause this phase jump.

Figure \ref{fig:par} shows that the long-term evolution of the unabsorbed flux, burst rate, and photon index of PL, with the strong correlations between each parameter.
Around the phase jump, a brief high burst rate interval appears at the peak of the flare, and the phase jump occurs just before this interval, which seems to suggest a connection between the phase jump and the X-ray burst.
The high burst rate interval accompanied by the enhancement of persistent emission has been observed in previous studies \citep[e.g.][]{2020ApJ...904L..21Y, 2022MNRAS.516..602B}, and the spectrum becomes harder and the photon index decreases \citep{2009ApJ...696L..74M}, which are consistent with the results in this study.
The burst rate rapidly decreases to zero within a few hours after the flare, and almost no bursts are detected thereafter.
The connections with the bursts and the hardness are related to the evolution of a twisted magnetic field \citep{2002ApJ...574..332T, 2010MNRAS.408.1387I}.
After the flare, the photon index tends to recover to the pre-flare level, the recovery process is relatively slow and does not reach the previous level in a short period of days.
This may indicate a certain degree of lag in the evolution of the system \citep{2007ApJ...654..470W}.
The simultaneous occurrence of phase jump, glitch, high burst rate interval, and the spectral evolution may indicate more complex scenarios, which may be caused by the mutation in twisted magnetic fields. 

We also noted the appearance of FRB 221014 just behind the flare as marked with the black dash-dotted line in Figure \ref{fig:par}, and this temporal consistency indicated a potential physical connection between FRBs and the twisted magnetic fields.
However, The triggering mechanism of FRBs is still unclear.
It is generally suggested that FRBs are generated by the magnetospheric activity of the magnetars, including internal \citep[e.g.][]{2018ApJ...868...31Y, 2020MNRAS.498.1397L} and external \citep[e.g.][]{2017ApJ...836L..32Z, 2020ApJ...897L..40D} triggering.
From the study of FRB 200428, FRBs preferably occur after the most active episodes triggered by giant glitches, which points to an active magnetic field, as the magnetic field rearrangement is completed, magnetic activity becomes less frequent and FRBs become more difficult to generate \citep{2024RAA....24a5016G}.
This supports the conjecture that a significant change in twisted magnetic fields leads to the phase jump, while the change in the magnetospheric structure results in the gradual appearance of the high burst rate interval and the FRB.
Following FRB 200428, several radio bursts emerge at different phases without significant X-ray variability \citep{2020ApJ...904L..21Y}. 
Similarly, subsequent to FRB 221014, there are also several radio bursts detected, with no significant changes observed in the timing or spectral properties of the persistent X-ray emission near FRB 221021 and FRB 221201 \citep{2024RAA....24a5016G}.
The outburst decays by $\gtrsim 50\%$, and lower data quality may make small changes invisible, which also indicates the diversity and potential differences of FRBs.

\begin{acknowledgments}
This work has made use of data from the NICER mission, and data and/or software provided by the High Energy Astrophysics Science Archive Research Center (HEASARC), a service of the Astrophysics Science Division at NASA/GSFC. 
This work is supported by the National Key R\&D Program of China (2021YFA0718500) from the Minister of Science and Technology of China (MOST). The authors thank supports from the National Natural Science Foundation of China under Grants U2038103, U2038101, U2038102 and 12373051. This work is also supported by International Partnership Program of Chinese Academy of Sciences (Grant No.113111KYSB20190020).
\end{acknowledgments}

%

\vspace{5mm}


\software{ASTROPY
\citep{2013A&A...558A..33A,2018AJ....156..123A},  
          XSPEC \citep{1996ASPC..101...17A}, and  STINGRAY \citep{2019ApJ...881...39H}.
          }




\bibliography{sample631}{}
\bibliographystyle{aasjournal}



\end{document}